\documentstyle[12pt,aps]{revtex}
\begin{document}
\title{
Chiral NN interactions in nuclear matter\bigskip}
\author{Boris Krippa \thanks{on leave from the 
Institute for Nuclear Research of
 the Russian Academy of Sciences, Moscow Region 117312,
Russia.}} 
\address{ Special Research Centre for Subatomic Structure of Matter,
Department of Physics and Mathematical Physics, The University
of Adelaide,  Australia 5005.\\}
\maketitle
\bigskip
\vspace{-5.5cm}
\hfill ADP-98-81/T347
\vspace{5.5cm}
\begin{abstract}
 We consider an effective field theory of NN system
in nuclear medium. The shallow bound states, which complicate the 
effective field theory analysis and lead to the large scattering length
in the vacuum case
do not exist in matter. We study
whether the chiral expansion of the effective potential can
be truncated in such situation. The cutoff regularization is used to render
the solution of the Bethe-Goldstone equation finite. We show that  
the next-to-leading order terms in the chiral expansion of the 
effective NN potential can indeed be interpreted as corrections so that
the truncation of the expansion is justified. It is pointed out hovewer
that it is still useful to treat the problem nonperturbatively since
it may allow the consideration of the nuclear systems with the density
smaller that the normal nuclear matter one. The possible directions of 
constructing the chiral theory of NN interaction in medium are outlined.

\end{abstract}

\vfill

\newpage

Effective Field Theory (EFT)
is becoming increasingly popular tool for studying nuclear interactions.
EFT is based on the idea of using in the low-energy region the Lagrangian
 with the appropriate effective degrees of freedom instead
of the fundamental ones (for review of EFT see,
for example \cite {Ma95}). This Lagrangian should include all possible terms 
allowed by the symmetries of the underlying QCD. The short-range physics 
is supposed to be approximated by the contact interactions. The states ,which
 can  be treated as heavy compared to the typical energy scale involved,
are integrated out and manifests themselves in the Low Energy Effective
Constants (LEC's). The physical amplitudes can be represented as the sum 
of certain graphs, each of them being of a given order in $Q/\Lambda$,
where $Q$ is a typical momentum scale and $\Lambda$ is a parameter 
close to the mass  of the nearest ``heavy'' particle which was integrated
 out. The relative contribution of each graph can be roughly estimated
using chiral counting rules \cite {We79}. The relevant degrees of freedom in
the nuclear domain are nucleons and pions. In the cases
 of meson-meson \cite {GL} and meson-nucleon\cite {Ec}
interactions the perturbative chiral expansion can be organized in 
consistent way. However, being applied to the NN system EFT encounters
serious problem which is due to the existence of bound or virtually bound
states near threshold \cite {We91}. It results in the large nucleon-nucleon
scattering length and makes the perturbative expansion divergent
so that chiral counting rules 
cannot be straightforwardly  transformed into the corresponding counting 
for the whole amplitude. Weinberg 
suggested \cite {We91} to apply chiral counting rules to the certain class 
of the irreducible diagrams which were then summed up to infinite 
order by solving the Lippmann-Schwinger (LS) equation. 
The irreducible diagrams can be treated as the effective potential
 in this case. However, being dimensionally regulirized,
such an effective theory exhibits rather small radius of
 convergence \cite {Ka96}
making the practical applications of the approach to the realistic case of 
NN interactions highly problematic if not impossible.\\
The modified formulation of EFT for NN problem, overcoming the 
abovementioned difficulties has been suggested in 
Refs. \cite {Ka98,VK98}. According to the strategy, adopted in
 \cite {Ka98,VK98} one needs to sum up some subclass of
 diagrams leading to the amplitude with fixed chiral dimension.
The contributions of the other graphs can be calculated perturbatively.
Different aspects of of the chiral NN problem were also considered
 in Refs. \cite {Fu,Co97}. In parallel with the 
development of the chiral theory for NN interactions the concept of EFT
has intensively been used to study the nuclear matter 
\cite {Se97,Ly,Lu,Fr}. In \cite {Se97} the effective
 chiral Lagrangian was constructed and the ``naturalness'' of the
 effective coupling constants has been demonstrated. The possible
counting rules for nuclear matter have been suggested in \cite {Lu}.
These two lines of development of the chiral nuclear physics are in some sense
similar to tendencies existed some time ago in the conventional nuclear
physics with phenomenological two body forces. On one hand, phenomenological
NN potentials had been using to describe nucleon-nucleon cross sections
and phase shifts. On the other hand, the nuclear mean field approaches
 had been proven successful in describing the bulk properties of nuclear matter.
It was therefore quite natural to consider the problem of NN 
interactions being put in nuclear medium, when two interacting particles
are moving in nuclear mean field. It then led to the famous
Bethe-Goldstone (BG) equation \cite {Be} for the G-matrix which is an 
analog of  scattering T-matrix, satisfying the LS equation.  
It seems therefore reasonable to follow the same strategy and, being
 equipped with
the chiral theory of NN interaction in vacuum, try to construct the chiral
G-matrix, describing the effective interactions of two nucleons in medium.
One can easily see the qualitative difference between vacuum and medium cases.
 In nuclear medium because of Pauli blocking  the intermediate 
states with the momenta
less than Fermi momentum $p_F$ are forbidden. Therefore, the nucleon propagator
does not exhibit a pole. Moreover, the shallow bound or virtual NN states,
which constitute the main difficulty of the problem in vacuum, simply do not
exist in nuclear matter because of the interaction of NN pair with nuclear mean
 field. It means that the effective scattering length becomes considerably
smaller compared to the vacuum one. It in turn would indicate that some variant
of chiral expansion may turn out possible. 
 
We start from the standard nucleon-nucleon effective chiral Lagrangian
which can be written as follows
\begin{equation}
{\cal L}=N^\dagger i \partial_t N - N^\dagger \frac{\nabla^2}{2 M} N
- \frac{1}{2} C_0 (N^\dagger N)^2\\ 
-\frac{1}{2} C_2 (N^\dagger \nabla^2 N) (N^\dagger N) + h.c. + \ldots.
\label{eq:lag}
\end{equation}
We consider the simplest case of the $^{1}S_0$ NN scattering and assume
zero total 3-momentum of NN pair in the medium. The inclusion of
the nonzero total 3-momentum does not not really change anything 
qualitatively and only makes the calculations technically more involved.   
The G-matrix is given by
\begin{equation}
G(p',p)=V(p',p) + M \int \frac{dq q^2}{2 \pi^2} \, V(p',q) 
\frac{\theta(q-p_F)}{M(\epsilon_{1}(p) +\epsilon_{2}(p')) - q^2} G(q,p),
\label{eq:LSE2}
\end{equation}
Here $\epsilon_1$ and $\epsilon_2$ are the single-particle 
energies of the bound nucleons.
They in principle are affected by the nuclear mean field but in this paper
we neglect nucleon self-energy
corrections in medium. The 
usually accepted strategy in  treating the chiral NN problem in vacuum
is the following. One calculates amplitudes up to  given chiral order
in the terms of the effective constants $C_0$ and $C_2$ which are then 
determined by comparing the obtained amplitude with the experimentally known
scattering length and effective radius. Having these constants determined one
can calculate the phase shifts and mixing parameters. In matter however
there is no such thing as phase shifts so we will proceed as follows.
We chose exactly solvable separable potential with parameters adjusted
to describe the experimental scattering length and effective radius
in vacuum. Then, using these parameters we calculate the in-medium effective
radius $r_m$ and scattering length $a_m$ by solving
 the BG equation.
Given $r_m$ and  $a_m$ we determine the effective constants $C_0$ and $C_2$.
The criteria of consistency we used is the 
size of the difference between $C_0$'s
calculated in the leading and subleading orders. In the vacuum case
the corresponding difference was found to be large \cite{Co97} indicating
 that the straightforward use  of 
chiral expansion of the effective NN Lagrangian when the 
series is truncated at a given order to get an effective potential to be
substituted to the LS equation  may not be a fully consistent procedure
and more subtle approaches (which were
recently suggested in Refs. \cite {Ka98,VK98} need to be 
developed to treat NN system by the EFT methods in completely consistent
 manner. 
As we already mentioned the situation is rather different in matter.
Let's first consider a simple separable potential
\begin{equation}
V=-\lambda |\eta\rangle\langle\eta|
\end{equation}
with the form factors

\begin{equation}
\eta(p)=\frac{1}{(p^2 + \beta^2)^{1/2}}
\end{equation}
 The solution of the LS equation is easily found to be
\begin{equation}
\frac{1}{T(k,k)}=
{V(k,k)^{-1}}\left[1 - M\int \frac{dqq^2}{4 \pi^{2}} \, \frac{V(q,q)}
{{k^2}- q^2}\right] 
\label{eq:sep}
\end{equation}
The experimental values of scattering length $a$ and effective radius $r$
are
\begin{equation}
a=-23.71\pm 0.013\, {\rm fm}\qquad {r_e}=2.73\pm 0.03\, {\rm fm}.
\label{eq:exp}
\end{equation}
These values can be reproduced if we chose
\begin{equation}
\lambda = 1.92\,\qquad \beta = 0.77\, {\rm fm}
\end{equation}
The solution of the BG equation for the separable potential is a simple
generalization of the one for LS equation 

\begin{equation}
G(k,k)=-
\eta^{2}(k,k)\left[\lambda^{-1}+ 
\frac{M}{2 \pi^{2}} \int{dqq^2} \, \frac{\theta(q-p_F) \eta^{2}(q,q)}{{k^2}- q^2}\right]^{-1}
\end{equation}
The parameters $\beta$ and $\lambda$ 
being substituted to the solution of the BG equation results in the 
in-medium scattering length $a_m$=-1.1 ${\rm fm}$ and effective
 radius $r_m$=1.8 ${\rm fm}$. One notes that effective radius is much less affected by the
medium effects. It is quite natural since the value of the effective radius is only
weakly sensitive to the bound state at threshold and  is of the ``natural'' size
already in the vacuum case. 
Comparing the vacuum and in-medium scattering lengths one can see
that the absolute value of the  in-medium scattering length is greatly reduced 
compared to the vacuum one. It clearly indicates that, as expected, the 
shallow virtual nucleon-nucleon bound state is no longer present in nuclear medium
and one can avoid significant part of the difficulties typical for the chiral NN
problem in vacuum. Having determined the values of scattering length and effective 
radius in medium one can now solve the BG equation using leading and sub-leading
orders of the NN effective chiral Lagrangian. The solution is quite analogous to 
the vacuum case \cite{Co97} and can be represented as follows  

\begin{equation}
\frac{1}{G(k,p_F)}=\frac{(C_2 I_3(k,p_F) -1)^2}{C_0 + C_2^2 I_5(k,p_F)
 + {k^2} C_2 (2 - C_2 I_3(k,p_F))} - I(k,p_F),
\label{eq:Tonexp}
\end{equation}
where we defined

\begin{equation}
I_n \equiv -\frac{M}{(2 \pi)^2} \int dq q^{n-1}\theta(q-p_F)\theta(\Lambda-q).
\label{In}
\end{equation}
 
and  

\begin{equation}
I(k) \equiv \frac{M}{2 \pi^{2}}\int dq  \, \frac{q^2\theta(q-p_F)\theta(\Lambda-q)}
{{k^2}- {q^2}}.\label{eq:IEdef}
\end{equation}
 We used the values $p_F$= 1.37 ${\rm fm^{-1}}$ and  $\Lambda$= 1.85 ${\rm fm^{-1}}$
for the Fermi-momentum and cutoff parameter respectively. The value of the cutoff
chosen deserves some comments. Our effective description is supposed to be 
valid in the region lying somewhat below the mass of the nearest ``heavy particle''.
In the other words it is natural  to be slightly below below the scale starting from which 
short-range degrees of freedom cannot safely be ignored in the effective Lagrangian. This 
scale approximately corresponds to the scale of short-range correlations, that is,
 2.2 - 2.3 ${\rm fm^{-1}}$. The description of the short-range correlations is hardly
 possible 
in the framework of the EFT so by choosing the value   $\Lambda$= 1.85 ${\rm fm^{-1}}$ we
hopefully retain only long-ranged effects. It is clear that pions should, in principle, be 
included explicitly in the theory. In this paper, which is the first attempt to consider
the possibility of constructing the effective chiral theory for the NN system in medium,
we do not include pionic effects. These effects cannot lead to the significant increase
of scattering length which is determined by the existence or absence the shallow bound state.
One notes, however, that taking into account nuclear pions may become important when 
the chiral G-matrix is used to describe the bulk properties of nuclear matter. This issue will
 be  addressed in the forthcoming paper.

Expanding the G-matrix in the powers of some external momentum $k$ and equating the 
corresponding coefficients in the chiral and effective range expansions one can get
the following expressions
\begin{equation}
\frac{M}{4 \pi a_m} = \frac{(C_2 I_3 -1)^2}{C_0 + C_2^2 I_5} +
 \frac{M}{2 \pi^2}(\Lambda - p_F),
\label{eq:C_0beta}
\end{equation}

\begin{equation}
\frac{M r_m}{8 \pi} =
\frac{ C_2 (2 - C_2 I_3) (C_2 I_3 - 1)^2}{(C_0 + C^2_2 I_5)^2 } +
 \frac{M}{2 \pi^2}\left[\frac{1}{\Lambda} - \frac{1}{p_F}\right]
\label{eq:C2beta}
\end{equation}
for the in-medium scattering length and effective radius respectively.
Considering  $a_m$ and $r_m$ as ``observables'' we demand that these
 quantities should be the same in the leading and subleading order. 
In vacuum case it resulted in significant difference between the values of the
parameter $C_0$ needed to fit the experimental values in leading and
subleading order indicating the potential problem with the straightforward
truncation of the chiral expansion of the effective NN potential. The same
 procedure leads in the in-medium case to the quadratic equation for the 
effective constant $C_2$ which has two solutions. One of them still results
in quite significant (although much less than in vacuum case) difference 
between $C_0$'s calculated in leading and subleading order, while the 
second one ( $C_2 \sim 0.63$)
gives rise to the approximately 40$\%$ correction, that is
\begin{equation}
 C_0^{(1)} \sim (0.58 - 0.6)  C_0^{(0)}
\end{equation}
Here $C_0^{(0)}$ and $C_0^{(1)}$ are the numerical values of 
the effective constant $C_0$ calculated in the leading and subleading order
respectively. The issue of interpretation of the first solution remains
open. It is probable, hovewer, that  using
the  experimental data on the nuclear matter binding energy may help to
exclude this solution as unphysical. Since the shallow bound state no longer
presents in the problem it looks more justified to use the solution leading
to the changes of $C_0$ which can really be viewed as
 a genuine ``correction''. One also notes that the value $C_2 \sim 0.63$
is  more suitable from the ``naturalness'' assumption point of view which seems
to hold in nuclear matter \cite{Se97}. 
 It is worth mentioning that
 in practical calculations one always needs to carry out the renormalization 
program and replace the bare parameters with the renormalized ones. 
The expressions relating the bare and renormalized couplings are,
in general highly nonlinear \cite{VK98}. In the leading order however they
coincide with each other so that the relatively moderate difference
between the bare  $C_0^{(0)}$ and $C_0^{(1)}$ effective coupling constants,
which can be treated as a correction,
should qualitatively reflect the similar behavior of the renormalized
ones. The fact of moderate changes in the value  $C_0$, when
  the next-to-leading
order terms are taken into account,  paves  the way for
the perturbative calculations, at least in principle. Hovewer, in spite of
this it is still 
useful to treat this problem in the nonperturbative manner. There are few 
reasons for the nonperturbative treatment. Firstly, the corrections themselves
are quite significant. Secondly, treating the problem nonperturbatively
may allow to consider the case of the densities smaller than the normal
nuclear one, where the nonperturbative aspects of the problem may play
 more important role.
 One notes however that BG equation is not supposed to be used
for vanishingly small densities so that the variations of density should
better be in the vicinity of the normal nuclear one. Thirdly, in the processed
involving both the nonzero density and temperature, such as heavy ion
collisions, the value of the Fermi-momentum can effectively be lowered again
making the nonperturbative treatment preferable.

In this, rather qualitative study we considered only one aspect of chiral
theory of the NN interactions in the nuclear medium. Many things remain to
be done in order to make the quantitative description of nuclear properties
possible. For example, one surely needs to include nuclear pions in the
 chirally symmetric manner and carry out the whole program of renormalization
like that recently presented for the NN interaction in vacuum \cite{VK98,Ka98}.
The corresponding LEC's in different spin-isospin channels should be extracted
from the experimental data available. Having these parameters fixed one
could determine the effective nuclear mean field which then can be used
to develop something like self-consistent chiral Hartree-Fock approach.

Taking into account that the shallow NN bound state
, the  main difficulty of the chiral theory 
of NN interaction in vacuum, does not exist in nuclear matter we remain 
cautiously optimistic about the perspectives of the chiral theory of nucleon-nucleon
interactions in nuclear medium.

\section*{Acknowledgments}

Author is very grateful for the support from SRCSSM where the main part of
 this work was done.

\end{document}